\documentclass[11pt,preprint]{aastex}
\usepackage{psfig}

\newcommand{\be}{\begin{equation}}
\newcommand{\ba}{\begin{eqnarray}}
\newcommand{\ee}{\end{equation}}
\newcommand{\ea}{\end{eqnarray}}  
\newcommand{\etal}{et al.\ }
\def\be{\begin{equation}} \def\ee{\end{equation}}      
  
\def\etal{{\it et al.~}}   
 
\def\HI{\hbox{H~$\scriptstyle\rm I\ $}}

\def\CIV{\hbox{C~$\scriptstyle\rm IV\ $}}

\def\OVI{\hbox{O~$\scriptstyle\rm VI\ $}} \def\nHI{{\rm HI}}
\def\nCIV{{\rm CIV}} \def\nOVI{{\rm OVI}}

\def\gsim{\lower.5ex\hbox{\gtsima}}
\def\lsim{\lower.5ex\hbox{\ltsima}} \def\gtsima{$\; \buildrel > \over
\sim \;$} \def\ltsima{$\; \buildrel < \over \sim \;$} \def\prosima{$\;
\buildrel \propto \over \sim \;$} \def\gsim{\lower.5ex\hbox{\gtsima}}
\def\lsim{\lower.5ex\hbox{\ltsima}}
\def\simgt{\lower.5ex\hbox{\gtsima}}
\def\simlt{\lower.5ex\hbox{\ltsima}}
\def\simpr{\lower.5ex\hbox{\prosima}} \def\la{\lsim} \def\ga{\gsim}
\def\ie{{\frenchspacing\it i.e. }} 

\def\gtsima{$\; \buildrel > \over \sim \;$}
\def\ltsima{$\; \buildrel < \over \sim \;$}
\def\gsim{\lower.5ex\hbox{\gtsima}}
\def\lsim{\lower.5ex\hbox{\ltsima}}
\def\simgt{\lower.5ex\hbox{\gtsima}}
\def\simlt{\lower.5ex\hbox{\ltsima}}
\def\simpr{\lower.5ex\hbox{\prosima}}
\def\la{\lsim}
\def\ga{\gsim}

\def\ie{{\frenchspacing\it i.e., }}

\def\E3{{\cal E}_{\rm g}^{III}}

\def\ozs{\Omega_Z^{sfh}}
\def\ozo{\Omega_Z^{obs}}

\begin{document}
\title{Where are the missing cosmic metals ?} 
\author{Andrea Ferrara\altaffilmark{1}, Evan Scannapieco\altaffilmark{2} \& Jacqueline Bergeron\altaffilmark{3}} 
\altaffiltext{1}{SISSA/International School for Advanced Studies, Via 
Beirut 4, I-34914, Trieste, Italy.}
\altaffiltext{2}{ Kavli Institute for Theoretical Physics,
Kohn Hall, UC Santa Barbara, Santa Barbara, CA 93106.}
\altaffiltext{3}{Institut d'Astrophysique de Paris - CNRS, 98bis Boulevard
Arago, 75014 Paris, France.} 

\begin{abstract}
The majority of the heavy elements produced by stars 2 billion years after the Big Bang (redshift $z\approx 3$) are presently undetected at 
those epochs. We propose a solution to this cosmic `missing metals' problem in which such elements are stored in 
gaseous halos produced by supernova explosions around star-forming galaxies. By using data from the ESO/VLT Large Program, 
we find that: (i) only 5\%-9\% of the produced metals reside in the cold phase, the rest being found in the hot  
($T=10^{5.8-6.4}$~K) phase; (ii) 1\%-6\% (3\%-30\%) of the observed \CIV (\OVI) is in the hot phase. We conclude that at $z \simgt 3$ more than 90\% of the metals 
produced during the star forming history can be placed in a hot phase of the IGM, without violating any observational 
constraint.  The observed galaxy mass-metallicity relation, and the intergalactic medium and intracluster medium 
metallicity evolution are also naturally explained by this hypothesis.
\end{abstract}


\section{Motivation}

In its original formulation (Pettini 1999), the `missing metals' problem was stated as follows. Studies of the comoving luminosity density of distant galaxies allow us to trace the cosmic star formation density (or history, SFH), $\dot\rho_\star(z)$, up to redshift $z_{max} \approx 7$. Assuming an initial  mass function of such stars (IMF), one can compute the specific fraction of heavy elements (`metals') they produce, $y$, and derive the metal production rate $\dot\rho_Z(z) = y \dot\rho_\star(z)$, whose integral from $z_{max}$ gives the density of cosmic metals in units of the critical density, $\ozs$, at any given $z$.   Early searches in cosmic structures for which the metal/baryon mass ratio\footnote{When necessary, we use the following cosmological parameters ($\Omega_\Lambda, \Omega_m, \Omega_b, n, \sigma_8, h$) = ($0.7, 0.3,
0.044, 1, 0.9,0.71$), consistent with {\it WMAP} results (Spergel et al. 2003), a solar metallicity $Z_\odot=0.0189$ by mass, and adopt the notation $Y_x=Y/10^x$} (metallicity,
$Z=\Omega_Z/\Omega_b$) can be derived either via intergalactic gas quasar absorption line experiments (Damped Ly$\alpha$ Absorbers [DLAs]
or the Ly$\alpha$ `forest') or through direct spectroscopic studies (Lyman Break Galaxies [LBGs]) have found that only $\ozo \la 0.20\ozs$ is stored in these components, \ie the large majority of the metals are `missing'. An analogous missing metal problem is also found by considering in a self-consistent manner the star formation rates and metallicities of DLAs alone (Wolfe et al. 2003, Prochaska et al 2003). 
Newly available high-quality data allow a more precise analysis of the problem. \\

\section{Stating the problem}
To re-evaluate $\ozs$ we use the most recent SFH compilation (Bouwens et al. 2004)  corrected upwards for the effects of dust obscuration by the prescribed (Reddy \& Steidel 2004) value of 4.5 at $z \ga 3$ and adopt $y=1/42$. Integration of $\dot\rho_Z(z)$ to $z=2.3$ yields $\ozs = 1.84 \pm 0.34\times 10^{-5}$.
Where should we look for these metals ?\\
The most obvious locations are the galaxies used to derive the SFH, \ie LBGs. These are characterized (Reddy \& Steidel 2004) by a mean comoving number density of  $6\times 10^{-3} h^3$~Mpc$^{-3}$  and  $Z=0.6 Z_\odot = 0.0113$.
Stellar masses can be constrained only by assuming a range of star formation histories of the form $SFR(t) \propto \exp(-t/\tau)$ and therefore they are somewhat uncertain. 
According to Shapley et al. 2005, they should be in the range $0.6 - 6 \times 10^{10} M_\odot$. Assuming the best fit
value $M_{\star}=2\times 10^{10} M_\odot$, we get $\Omega_Z^{lbg} = 3.4 \times 10^{-6} M_{\star,10} \approx 0.18 \ozs$. 
If metals are not in LBG stars or gas, they could be in DLAs or the IGM.
The metal content of DLAs is derived by noting that (Rao \& Turnsheck 2000, Prochaska \& Wolfe 2000)  at $z\approx 3$ their neutral ($\approx$ total) gas density $\Omega_g^{dla}=10^{-3}$ and metallicity $Z=3.8\times 10^{-4}$ combine to give  $\Omega_Z^{dla} = 3.8 \times 10^{-7} < \ozo \ll \ozs$; we can therefore neglect their contribution. In the following, we 
correct for LBG contribution by (re)defining the cosmic density of missing metals $\ozs \equiv \ozs - \Omega_Z^{lbg}$.\\
Hence, the missing metals should be found outside the main body of galaxies or DLAs. There are essentially two possibilities: (a) they could reside in the Ly$\alpha$ forest, or (ii) in the halos of the galaxies that have produced and ejected them. Note that the distinction between these two components is somewhat ambiguous. Our working definition is that galactic halos are gravitationally bound structures around galaxies; they are special sites affected by galactic winds.  

The most widely studied tracers of Ly$\alpha$ forest metallicity are  \CIV and \OVI absorption lines.
The fraction of the critical density $\rho_{c}$ contributed by a given
element ion, $E_i$, of mass $m_E$ residing in the Ly$\alpha$ forest is
given by
\begin{equation}
\Omega_{E_i}^{ly\alpha}= {H_0\over \rho_{c} c} {\sum N_{E_i}\over \sum
\Delta X}m_E
\label{omegaei}
\end{equation}
where $\Delta X(z_+, z_-)= \int_{z_-}^{z_+} dz (1+z)^2 E(z)^{-1}$ is the absorption distance (Bahcall \& Peebles 1969),
with $E(z)=[\Omega_\Lambda + \Omega_m (1+z)^3]^{1/2}$; sums are
performed over all the redshift intervals $z_- < z  < z_+$ over which
an ion column density $N_{E_i}$ is detected.
To determine $\Omega_{\nOVI}^{ly\alpha}$ and $\Omega_{\nCIV}^{ly\alpha}$, we use data from the ESO VLT Large Program\footnote{http://www2.iap.fr/users/aracil/LP/} (Bergeron et al. 2002) which provides high S/N,
high resolution spectra of an homogeneous sample of 19 QSOs in $1.7 < z < 3.8$;  $\Omega_{\nOVI}^{ly\alpha}$ is currently available for four LP lines of sight (Bergeron et al. 2002, Bergeron \& Herbert-Fort 2005), for which we find $\Omega_{\nOVI}^{ly\alpha}= 1.3\times 10^{-7}$ ; two other recent estimates (Simcoe, Sargent \& Rauch 2004, Carswell, Schaye \& Kim 2002), give $\Omega_{\nOVI}^{ly\alpha}=(1.1, 0.9) \times 10^{-7}$. We adopt the sightline--weighted mean of the three values allowing for the largest error, $\Omega_{\nOVI}^{ly\alpha}=1.1\pm 0.3 \times 10^{-7}$.  From the \CIV  absorber distribution (Aracil et al. 2004, Scannapieco et al. 2005) in the column density range $12 < \log N_{\nCIV} < 16$ we obtain $\Omega_{\nCIV}^{ly\alpha}=7.5\pm 2.2 \times 10^{-8}$ (statistical error).  This value is about two times higher than previous determinations (Songaila 2001; Simcoe, Sargent \& Rauch 2004, Schaye et al 2003) which could not account for the contribution of strong ($\log \CIV > 14$) absorption systems.
Combining the average measured $N_{\nCIV}$--$N_{HI}$  and $N_{\nOVI}$--$N_{HI}$
correlations (Aracil et al. 2004) with
the measured distribution of weak $HI$ absorbers (Petitjean et al 1993),
we have checked that  systems  with $\log N_{\nCIV} < 12$  contribute
less than 1\%, well within the quoted error.
For a (meteoritic) solar carbon logarithmic abundance (in number) $A_C=8.52$ with respect to hydrogen ($A_H=12$),  we conclude that only a fraction
$\Omega_{\nCIV}^{ly\alpha}/\Omega_C^{sfh}= 2.4\times 10^{-2}$ of the produced carbon is observed in the \CIV state. Repeating the procedure
for O ($A_O=8.83$), gives a ratio $\Omega_{\nOVI}^{ly\alpha}/\Omega_O^{sfh}=1.3\times 10^{-2}$. To account for all uncertainties above, we will consider
values in the range $ 1.4 \times 10^{-2}  < \Omega_{\nCIV}^{ly\alpha}/\Omega_C^{sfh} <   4.0 \times 10^{-2}$ and
$ 8.1\times 10^{-3}   < \Omega_{\nOVI}^{ly\alpha}/\Omega_O^{sfh} <   2.1 \times 10^{-2}$.\\
We now determine the physical conditions of the gas hiding the missing C and O.  Numerical simulations (Dav\'e et al. 2001) suggest that the intergalactic medium [IGM] might be a two-phase system made by a cool ($T_c\approx 10^{4-4.5}$~K), photoionized phase, and a hot, collisionally ionized one. We impose 
the following conditions separately for each ion (\CIV,\OVI) and element (C,O): (1)  the observed ionic abundance is the sum of the
abundances in the two phases; (2) the SFH-produced element abundance is the sum of the element abundances in the two phases; (3) the elements are
in the same abundance ratios in the two phases.  More explicitly, these conditions can be mathematically expressed as
\ba                    
f_C^c \Omega_C^c + f_C^h \Omega_C^h &=& \Omega_{\nCIV}^{ly\alpha} \\   
f_O^c \Omega_O^c + f_O^h \Omega_O^h &=& \Omega_{\nOVI}^{ly\alpha} \\    
\Omega_C^c + \Omega_C^h &=& \Omega_{C}^{sfh} \\   
\Omega_O^c + \Omega_O^h &=& \Omega_{O}^{sfh} \\   
\Omega_C^c - A\Omega_O^c &=& 0 \\   
\Omega_C^h - A\Omega_O^h &=& 0 
\ea                 
After some simple algebra, the above equations reduce to:
\be
{{\Omega_{\nCIV}^{ly\alpha}/\Omega_C^{sfh} - f_C^h}\over {f_C^c -
f_C^h}} = {{\Omega_{\nOVI}^{ly\alpha}/\Omega_O^{sfh} - f_O^h}\over
{f_O^c - f_O^h}},
\label{rel}
\end{equation} 
where  $f_i^j \equiv f_i^j(\Delta_j, T_j, {\cal U}_j)$ is the ionization correction for the considered ion (\CIV or \OVI) of a
given element ($i=C,O$) in the cold or hot phase, ($j=c,h$), respectively; the overdensity, $\Delta_j$ and temperature, $T_j$, of
the two phases are the unknowns of the problem; finally, $A$ is the abundance ratio of the two elements.
We complement these conditions by further imposing that the pressure of the cool 
phase does not exceed that of the hot one and assuming a temperature-density relation for the cold phase  $T=T_0\Delta_c^\gamma$, 
(with $T_0=2\times 10^4$~K and $\gamma=0.3$),  as inferred from the Ly$\alpha$ forest data. The value of the photoionization parameter, ${\cal U}_j=n_\gamma/n_j$, is fixed by the
ionizing photon density $n_\gamma$ of the assumed UV background spectrum (Haardt \& Madau 1996) shifted so that the intensity at 1 Ryd is
$J_\nu= 0.3\times 10^{-21}$~erg ~s$^{-1}$~Hz$^{-1} = 0.3~J_{21}$, corresponding to a  hydrogen photoionization rate  $\Gamma=0.84\times
10^{-12} {\rm s}^{-1}=0.84~\Gamma_{12}$, in agreement with Bolton \etal 2005.  Finally, we warn that deviations from solar abundances might be possible, and indeed there are hints that oxygen 
might be  overabundant (Telfer et al. 2002; Bergeron et al. 2002); here we neglect this complication.

\section{A possible solution}
By solving eq. \ref{rel}, we obtain the results plotted in Fig. 1. The hot phase is characterized by a wide density range, $\log \Delta_h > 0.4 $ and a restricted temperature range $5.8 < \log T_h < 6.4$.  We find that: (i) only 5\%-9\% of the produced metals reside in the cold phase, the rest being found in the hot phase; (ii) 1\%-6\% (3\%-30\%) of the
observed \CIV (\OVI) is in the hot phase. We conclude that more than 90\% of the metals produced during the star forming history can be placed in a hot phase of the IGM,
without violating any observational constraint. To further constrain the hot phase parameter range, we have searched in the LP \CIV line list for components with large Doppler parameters. We find no lines with  $b_\nCIV \ge 26.5$~km~s$^{-1}$, corresponding to $\log T_h > 5.7$; this result seems to exclude the high density and temperature region of
the allowed parameter space in the middle panel of Fig.  1. We checked that the above findings are insensitive to variations of $\Gamma_{12}$ of $\pm 50\%$; however, \OVI/\CIV ratios in the cold phase might depend on the UVB shape around 4 Ryd.\\
The derived values of $T_h$ and $\Delta_h$ are suggestive of regions likely to be found around galaxies; moreover,  $10^6$~K gas temperature would have a 
scale height of $> 10$~kpc, hence it cannot be confined in the disk. To test this hypothesis we resort to cosmological simulations. As an illustration, Fig. 2 shows
the temperature and velocity structure in a 2D cut through the center of a simulated galaxy (we used the multiphase version
[Marri \& White 2003] of the GADGET2 code to 
simulate a comoving $10 h^{-1}$~Mpc$^3$ cosmic volume) at redshift $z=3.3$; its total (dark + baryonic) mass is $2\times 10^{11} M_\odot$, the star formation 
rate $\approx 20 M_\odot$~yr$^{-1}$. This galaxy has been selected to match LBG properties, but it  is not unusual in the simulation volume. As often observed in 
LBGs, a strong galactic wind is visible, whose expansion is counteracted by  energy losses due to cooling and gravity,  and ram pressure exerted by the infalling gas. 
Infall is particularly effective at confining the wind into a broadly spherical region of physical radius $\approx 300$~kpc, into which cold infalling
streams of gas penetrate. Inside such wind-driven bubble the temperature (Fig. 2)  is roughly constant $T\approx 10^6$~K, whereas the density spans values of  
$0  < \log \Delta < 5$ [$\Delta(z=3.3)=1$ corresponds to $\approx 2 \times 10^{-5}$~cm$^{-3}$]. The cool phase is evident in  the outer boundary
of the bubble, where cooling interfaces arise from the interaction with infalling streams.  Hence halos of LBGs seem to meet the requirements as repositories of 
missing metals.\\
Additional support for this conclusion comes from studies of the correlation properties of \CIV and \OVI absorbers (Pichon et al 2003, Aracil et al. 2004, Bergeron \& Herbert-Fort 2005), which conclude that: (i) \OVI
absorption in the lowest density gas is usually (about 2/3 of the times) located within $\approx 300-400$~km~s$^{-1}$ of strong \HI absorption lines; (ii) the \CIV correlation function is consistent 
with metals confined within bubbles of typical (comoving) radius  $\approx 1.4h^{-1}$ Mpc  in halos of mass $M \ge 5\times 10^{11} M_\odot$ at $z=3$.  
If each such objects hosts one bubble, the cosmic volume filling factor of metals is $f_Z =11\%$; it follows that halo metallicity is $\Omega_Z^{sfh}/f_Z \Omega_b=0.165 Z_\odot$. 
A  temperature of  $\log T_h =5.8$ corresponds to \HI (\OVI) Doppler parameters $b_\nHI =102$ ($b_\nOVI=25.5$)~km~s$^{-1}$
and to $N_\nOVI/N_\nHI=3$; absorbers with $\log N_\nOVI = 13$ are detectable for $b_\nOVI = 25.5$~km~s$^{-1}$
but the corresponding $\log N_\nHI = 12.4$ ones for $b_\nHI = 102$~km~s$^{-1}$ are not. This raises the possibility of finding \OVI
absorbers without associated \HI.\\
\section{Implications}
The scenario proposed leads to several interesting consequences. First, metals produced by LBGs do not seem to be able to escape from their
halos, due to the confining mechanisms mentioned above. This is consistent with the prediction (Ferrara, Pettini \& Shchekinov 2000) 
that galaxies of total mass ${\cal M} > 10^{12}(1+z)^{-3/2}M_\odot$ do not eject their metals into the IGM. Interestingly, the
metallicity-mass relation recently  derived from the SDSS (Tremonti et al. 2004) shows that galaxies with {\it stellar} masses above $3\times 10^{10} 
M_\odot$ (their total mass corresponds to ${\cal M}$ for  a star formation efficiency $f_\star = 0.2$) chemically evolve as ``closed boxes,'' \ie they retain their heavy
elements.
Second, the nearly constant ($2 \le z \le 5$, $Z \approx 3.5 \times 10^{-4} Z_\odot$) metallicity of the low column density IGM (Songaila 2001) is naturally explained by the  decreasing efficiency of metal loss from larger galaxies. Early pollution from low-mass galaxies allows a sufficient time for metals to cool after ejection; however, the majority of metals in LBG halos at lower redshifts are still too hot to be detected.  Hence their contribution to the metallicity evolution of the IGM cannot be identified by absorption line experiments, which mostly sample the cool phase of the forest.   
Third, the rapid deceleration of the wind results either in a quasi-hydrostatic halo or in a `galactic fountain' if radiative losses can cool the halo gas. In both cases this material is very poorly bound  and likely to be stripped by ram pressure if, as it seems reasonable, the galaxy will be incorporated in the potential well of a larger object (galaxy group or cluster) corresponding to the next level of the hierarchical structure growth. Turbulence and hydrodynamic instabilities associated with this process are then likely to efficiently mix the metals into the surrounding gas within approximately a sound crossing time of $\sim 1$~Gyr, or $\Delta z \approx 0.5$.  If metals produced and stored in LBG halos by $z=2.3$ end up in clusters, than the average
metallicity of the intracluster medium is  $Z_{ICM}= \ozs /\Omega_{ICM} = 0.31 Z_\odot$, having assumed (Fukugita, Hogan \& Peebles 1998) $\Omega_{ICM}=0.0026 h{^{-1.5}_{70}}$. Not only is this number tantalizingly close to the observed value at $z=1.2$ (Tozzi et al. 2003), but we also predict that little evolution  will be found in the ICM metallicity up to 
$z \approx 2$ as essentially  all the metals that could have escaped galaxies during cosmic evolution had already done so  by this epoch.

\acknowledgments 
We thank P. Rosati for discussions and
A. Fangano for help with data analysis.  We acknowledge partial
support from  the Research and Training Network `The Physics of the
Intergalactic Medium' set up by the European Community under the
contract HPRN-CT2000-00126 RG29185. This work has been completed
during the  KITP Program ``Galaxy-IGM Interactions'', funded by NSF
Grant PHY99-07949, whose support is  kindly acknowledged.

\fontsize{10}{10pt}\selectfont

\begin{figure}
\centerline{\psfig{figure=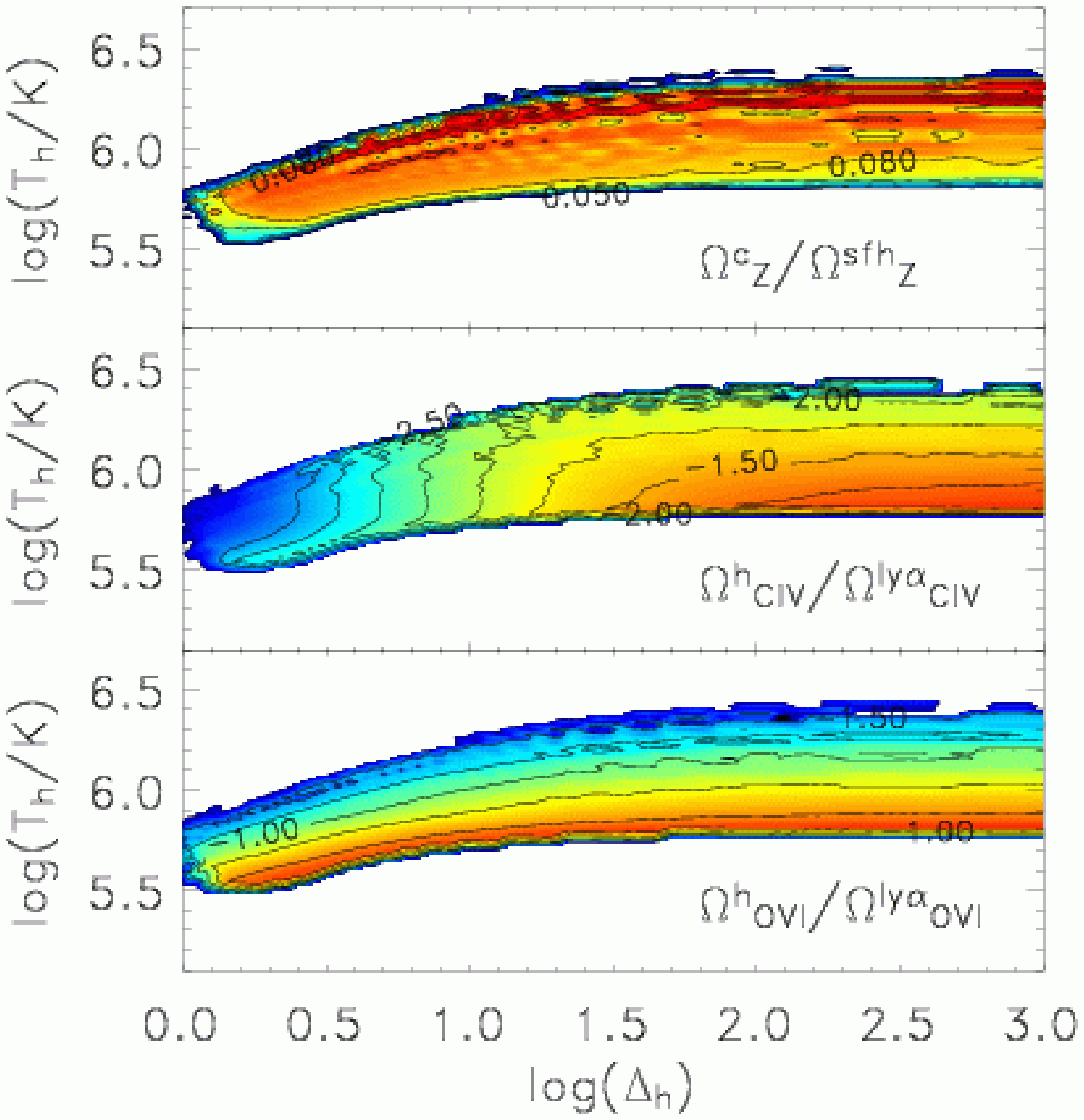,height=15cm}}
\caption{\footnotesize Allowed hot phase gas temperature, $T_h$, and
overdensity, $\Delta_h$, regions.  Also shown are the corresponding
isocontours of the $\Omega_Z^c/\Omega_Z^{sfh}$ ({\it upper panel}),
$\Omega_{\nCIV}^h/\Omega_{\nCIV}^{ly\alpha}$ ({\it middle panel})  and
$\Omega_{\nOVI}^h/\Omega_{\nOVI}^{ly\alpha}$ ({\it bottom panel}).}
\label{fig:1}
\end{figure}

\begin{figure}
\centerline{\psfig{figure=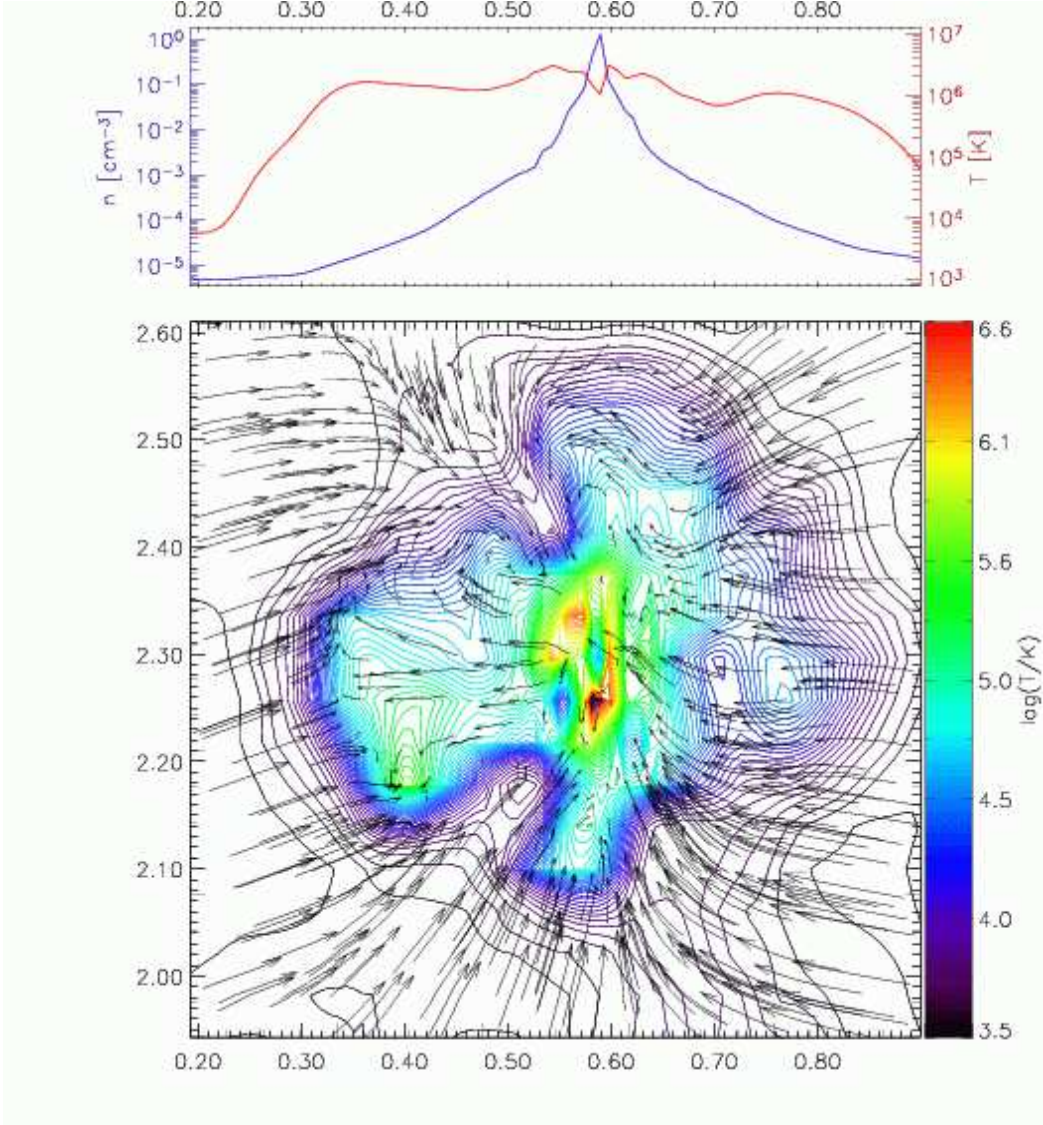,height=15cm}}
\caption{\footnotesize {\it Bottom panel:} Temperature map (physical
Mpc units) of a plane  through the center of a simulated starburst
galaxy at $z=3.3$ with properties typical of Lyman Break Galaxies;
longest velocity vectors correspond to $v=150$~km~s$^{-1}$. {\it Upper
panel:} 1D cuts parallel to the horizontal axis and passing through
the center of the map in the bottom panel  showing the gas density
(blue) and temperature (red)}
\label{fig:2}
\end{figure}
\end{document}